\documentclass[aps,pre,twocolumn,groupedaddress, showkeys, a4paper]{revtex4}
\usepackage{amsmath}
\usepackage{graphicx}
\usepackage{hyperref}

\begin{document}

\title{Jerarca: efficient analysis of complex networks using hierarchical clustering}

\author{Rodrigo Aldecoa}
\author{Ignacio Mar\'in}
\email[]{imarin@ibv.csic.es}
\affiliation{Instituto de Biomedicina de Valencia.
Consejo Superior de Investigaciones Científicas (IBV-CSIC)
Calle Jaime Roig 11. Valencia, Spain}


\begin{abstract}
\textbf{Background:} How to extract useful information from complex biological networks is a major goal in many fields, especially in genomics and proteomics. We have shown in several works that iterative hierarchical clustering, as implemented in the UVCluster program, is a powerful tool to analyze many of those networks. However, the amount of computation time required to perform UVCluster analyses imposed significant limitations to its use.

\textbf{Methodology/Principal Findings:} We describe the suite Jerarca, designed to efficiently convert networks of interacting units into dendrograms by means of iterative hierarchical clustering. Jerarca is divided into three main sections. First, weighted distances among units are computed using up to three different approaches: a more efficient version of UVCluster and two new, related algorithms called RCluster and SCluster. Second, Jerarca builds dendrograms based on those distances, using well-known phylogenetic algorithms, such as UPGMA or Neighbor-Joining. Finally, Jerarca provides optimal partitions of the trees using statistical criteria based on the distribution of intra- and intercluster connections. Outputs compatible with the phylogenetic software MEGA and the Cytoscape package are generated, allowing the results to be easily visualized.

\textbf{Conclusions/Significance:} The four main advantages of Jerarca respect to UVCluster are: 1) Improved speed of a novel UVCluster algorithm; 2) Additional, alternative strategies to perform iterative hierarchical clustering; 3) Automatic evaluation of the hierarchical trees to obtain optimal partitions; and, 4) Outputs compatible with popular software such as MEGA and Cytoscape.
\end{abstract}


\maketitle

\section*{INTRODUCTION}
There are many types of data, both biological and non-biological, which can be represented as undirected graphs. Examples in biology are networks based on protein-protein interaction data, those based on shared protein domains, genetic interaction networks or coexpression networks. Developing heuristic strategies to extract useful information from them is an active field of research (reviewed in \cite{1,2,3}). A typical problem is how to generate partitions of a network in order to establish clusters, groups of tightly connected units. There are two basic general strategies to perform such a task. One option is to search for densely connected modules, for instance using a local evaluation function that measures when adding or eliminating units leads to a significant decrease of the average density of connections within a group (see e. g. refs. \cite{4,5,6,7,8,9}). A second possibility is to generate complete partitions of the graph, assigning each unit to a cluster. This requires global parameters to evaluate the quality of the alternative partitions \cite{10,11,12}. Although both methods have advantages and drawbacks, the latter should be considered preferable on theoretical grounds, given that it allows classifying all the units of the network.

To classify data, hierarchical clustering has several advantages over other procedures. First, it is a fully unsupervised method. In the case of networks, this allows to cluster all units without having to specify a priori the number of clusters present. In addition, the generation of a hierarchical tree provides not only partitions of the network (either by how units are grouped in agglomerative clustering, or by how the units are divided into groups, in divisive clustering), but also allows to visualize how the basic, first-order clusters are combined into higher-level groups. However, the development of hierarchical clustering strategies to analyze networks is problematic. Particularly, clustering unweighted undirected graphs (e. g. networks of interacting units) is seriously hampered by the "ties in proximity" problem (discussed in \cite{12}). In this type of networks, the distance between two units is defined as the minimal number of edges that must be walked to connect them. Then, in typical biological networks – large and with small-world properties – the number of tied distances is astronomical. This makes it impossible to directly obtain a reasonable hierarchical tree based on the distances among units. The problem caused by the ties is that in each step of the clustering process a large number of alternative agglomerations (or divisions) are possible. Several authors attempted to solve this problem by using measures of proximity among units different from their distances \cite{13,14,15}. However, to justify the usage of any of these alternative parameters is difficult. A few years ago, we devised a valid strategy to solve the ties in proximity problem \cite{12}. The first step consists in generating a large number of alternative, mathematically equivalent partitions of the network using the distances among the units (\textit{primary distances}, according to our nomenclature) and conventional (e. g. average linkage) hierarchical clustering. The results are then averaged to obtain a weighted distance measure for each pair of units (\textit{secondary distance}). This distance corresponds to the fraction of alternative partitions in which two units are assigned to different clusters. Finally, a dendrogram is obtained from the matrix of secondary distances. This strategy, which we called iterative cluster analysis, has already empirically demonstrated its usefulness. High-quality dendrograms have been obtained from complex networks derived from different types of biological data \cite{16,17,18}. However, performing iterative hierarchical clustering has been so far hampered by the intrinsic slowness of obtaining a representative set of partitions. For example, our original program, UVCluster \cite{12}, runs in $O(n^3)$ time, $n$ being the number of nodes. For this reason, the largest analysis published so far corresponds to a network with just 632 units \cite{18}.

In this work, we describe a suite of programs called Jerarca (Spanish for hierarch), which contains new, efficient algorithms to perform iterative hierarchical cluster analyses. One of them is basically a faster implementation of the UVCluster program. The other two, RCluster and SCluster, provide alternative ways to obtain the matrices of secondary distances from a graph. In addition, for the conversion of the matrix of distances into a dendrogram, two well-known phylogenetic algorithms, UPGMA and Neighbor-Joining \cite{19,20,21} have been included in Jerarca. Finally, Jerarca also includes two different mathematical criteria to determine the best partition of the dendrogram into clusters. The first one is a parameter called modularity (Q) \cite{10}, which has been extensively used to measure community structure in networks. As an alternative, we include a modification of a hypergeometric distribution-based index suggested in one of our previous works \cite{12}. Several output files, useful to edit and visualize the results, are generated by the program. All these options make Jerarca much more efficient and versatile than our original UVCluster program.

\section*{METHODS}

\begin{figure*}[ht]
\includegraphics[scale=0.15]{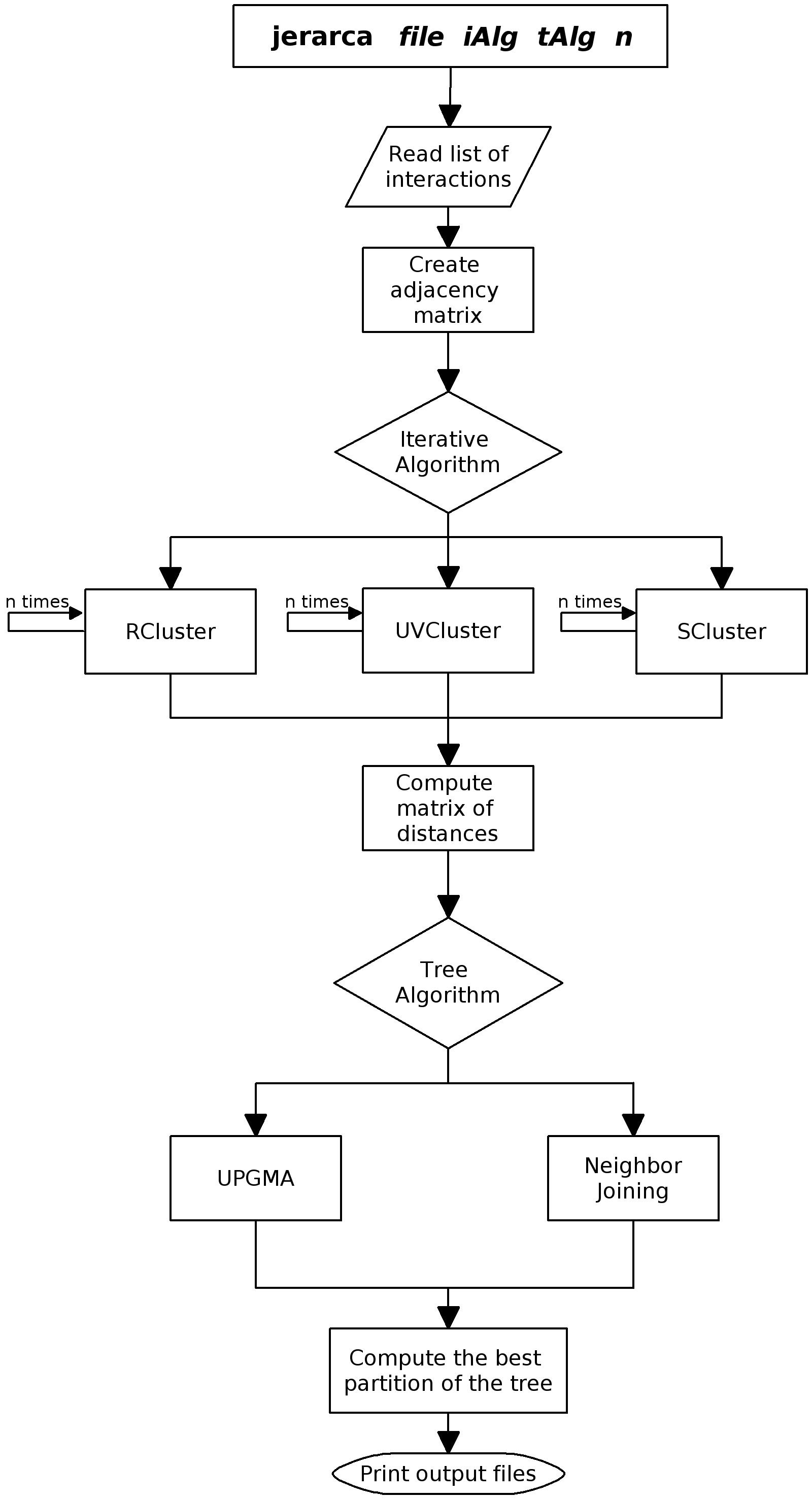}%
\caption{\label{fig:1} Control flowchart of Jerarca.
The four input parameters are file (list of interactions that represent the edges of the network), iAlg (iterative algorithm to use), tAlg (tree algorithm to use) and n (number of iterations to perform).}
\end{figure*}
The Jerarca suite has been written in \textit{C++}. Both the source code and compiled versions for Windows and Linux platforms are freely available at \href{http://jerarca.sourceforge.net}{http://jerarca.sourceforge.net}. Figure \ref{fig:1} details the control flow structure of the code. To perform a round of analyses, the user must execute the program from a command window, writing four parameters in the following order: 1) the name of a text file that describes the list of edges of the graph. The names of two linked nodes, separated by a tab or space, must be written in each line of the file; 2) the algorithm(s) chosen to iteratively calculate the matrix(ces) of secondary distances; 3) the algorithm(s) that will be used to obtain the dendrogram; and, 4) the number of iterations to be performed. Therefore, a typical Jerarca input has the following structure (parameters are indicated in brackets):

\textit{jerarca [Name of the file]}

\textit{[Iterative algorithm:(uv, r, s, all)]}

\textit{[Tree algorithm: (u, nj, all)]}

\textit{[Number of iterations]}

For the iterative algorithm, four options are valid: \textit{uv} (UVCluster), \textit{r} (RCluster), \textit{s} (SCluster) and \textit{all}. This last option will produce three parallel solutions, one for each available algorithm. For the tree algorithm, three options are valid: \textit{u} (UPGMA), \textit{nj} (Neighbor-joining) and \textit{all}. This last option again will produce two solutions, one for each algorithm.

\begin{figure*}[ht]
\includegraphics[scale=0.4]{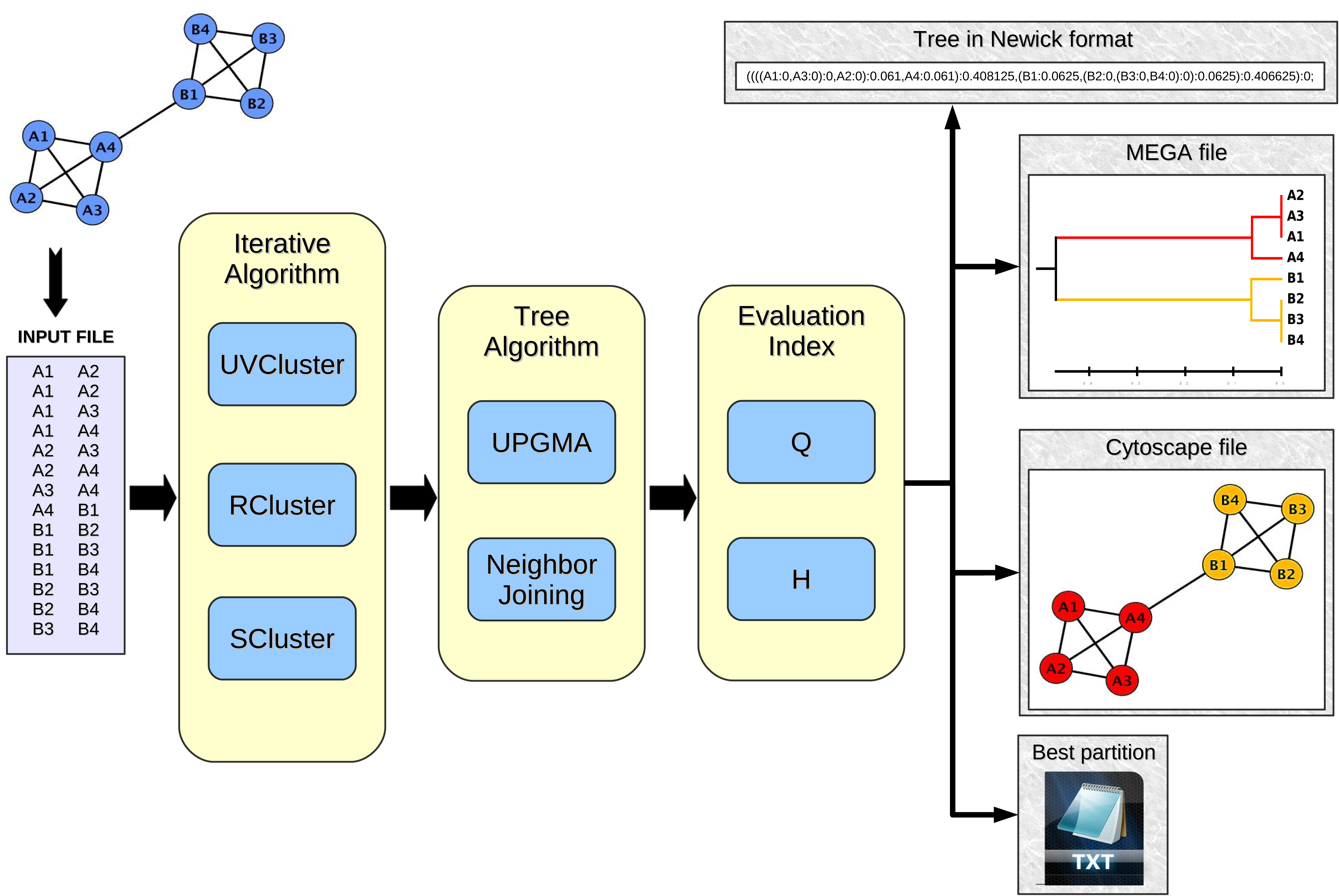}%
\caption{\label{fig:2} A typical analysis with Jerarca. The user specifies the input file where the graph is represented. It is analyzed by the program through diverse algorithms returning four different outputs: the tree in Newick format, a MEGA-compatible file, a file with attributes for Cytoscape and a text file containing the optimal partition of the tree.}
\end{figure*}

A typical Jerarca analysis is shown in Figure \ref{fig:2}. In summary, the program reads the input file and creates the adjacency matrix A of the graph: $A_{ij} = 1$ if vertices i and j are connected and $A_{ij} = 0$ otherwise. Then, it applies the iterative algorithm(s) selected as many times as the number of iterations specified. To calculate the matrix of secondary distances, the algorithm saves, for each pair of nodes, the number of iterations in which they have been clustered separately, and the secondary distances between each two units are calculated by dividing those values by the number of iterations. After creating the matrix of secondary distances, the program uses the phylogenetic algorithm(s) chosen to build a dendrogram. The program finally evaluates, using the two indices implemented, each level of the dendrogram and saves the optimal partition of the tree for each index (see below). Several convenient output files (described also in detail below) are generated.

\subsection*{Details of the iterative algorithms}
We recently developed several novel ideas that are the basis of Jerarca. We first thought a way to notably improve the speed of the UVCluster program. UVCluster contained a parameter called \textit{Affinity Coefficient (AC)}, which sets how permissive the clustering process is, in such a way that the lower the \textit{AC} value, the larger the average distances among clustered units can be (see \cite{12} for a detailed explanation). The maximum value of $AC = 100$ implies that only units that are directly connected in the graph are clustered together. Very significantly, this value was the only used in all our subsequent works \cite{16,17,18}. Not a single useful application for other values has ever been found. This has an important consequence, given that, if we fix $AC = 100$, UVCluster-based iterative hierarchical clustering can be performed using the adjacency matrix of the network instead of the matrix of primary distances. This avoids computing the primary distances among all units using Floyd's algorithm, whose time complexity is $O(n^3)$. Once noticed that important point, we decided to generate a new version of UVCluster implementing this new approach. It turns out that this improved version is qualitatively faster than our former program, running in $O(n^2)$ time.

\begin{figure*}[ht]
\includegraphics[scale=0.73]{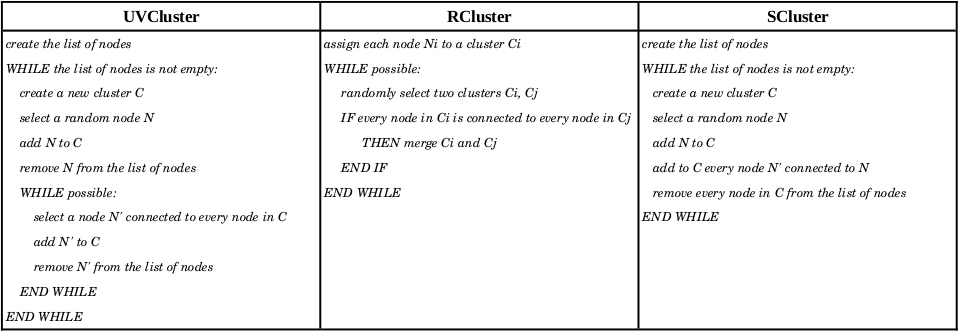}%
\caption{\label{fig:3}Main loop of the three iterative clustering algorithms implemented in Jerarca. An iteration defines a partition of the network by assigning the nodes to clusters. These loops are repeated as many times as iterations are specified by the user.}
\end{figure*}

Two new algorithms, called RCluster and SCluster, described here for the first time, provide alternative ways to establish the matrix of secondary distances, following strategies related to the one implemented in the new version of UVCluster. These programs use alternative methods to select the units to be merged. Figure \ref{fig:3} shows a compact, technical description of their differences. However, we think that the reader may benefit from the following verbal summary of how the three programs work. The differences in the clustering process are as follows:

\begin{enumerate}
\item To select which units to merge, UVCluster generates in each iteration a list in which the units are randomly ordered and then proceeds to generate a cluster taking the first unit in that list and searching for all the units that can be merged to that one, according to the provided \textit{AC} parameter. If $AC = 100$ (fixed value in the new version of the program) this means that UVCluster establishes cliques, i. e. groups in which each unit is connected with all the rest of units in the group. Once the largest clique that can be formed from the first selected unit is found, the units of that clique are set apart (i. e. they are considered to form a cluster) and the next unit still available in the list is used to start again the same process. This is a greedy algorithm, which tends to favor finding compact clusters.
\item Our second algorithm, RCluster (R meaning random), also establishes cliques but, instead of using a starting unit and greedily making a particular cluster to grow from it, RCluster in each step randomly merges two clusters, provided that all their units are connected (i. e. they form, after being merged, a clique). The program follows a hybrid strategy to select the clusters. To start with, the program simply randomly picks up two clusters, establishes whether they can be merged or not and, if indeed it is possible to merge them, puts all the units together into a single, new cluster. While there are many clusters that can be merged, this simple strategy is very efficient and it has the big advantage of not requiring to recalculate the adjacency matrix in each merging step, something that is very time consuming for large graphs. However, as the merging process progresses, the likelihood of finding mergeable clusters just by randomly picking up two of them gets smaller. It is then convenient to shift to a second strategy, which is indeed based on generating in each step of the merging process an adjacency matrix, in which a value $A_{ij} = 1$ means that the units of the two clusters (i, j) form a clique. This second strategy is implemented in two steps: 1) The program generates an adjacency matrix and then randomly searches for a $A_{ij} = 1$ value in that matrix to merge two clusters; 2) It recalculates the adjacency matrix. Logically, for the newly formed cluster, it assigns a value of “1” with another cluster only when all the units in both clusters are connected. These two processes are repeated until no clusters can be merged. The transition from the first to the second strategy occurs when $n$ random picks, $n$ being the number of nodes of the network, have failed to find two mergeable clusters. Empirical analyses have shown this to be a convenient cutoff. Notice that, in RCluster, and differently from what occurs in UVCluster, multiple clusters grow at the same time. However, the process of choosing a random pair of clusters to merge in each iteration makes the program slower than the current version of UVCluster. We found that it runs in $O(n^2 log n)$ time.
\item Finally, the third alternative is our novel SCluster algorithm (S stands for simple), which is both our greediest and our fastest algorithm, running in $O(n log n)$ time. SCluster just picks up a unit by random and then collapses in a cluster that unit with all the units directly connected to it. These units are removed from the graph and then another unit is randomly chosen and the process is repeated until no further units remain. Notice the difference with UVCluster and RCluster: the units collapsed in a cluster do not have to be all connected among them (forming cliques) but just linked to the initial unit.
\end{enumerate}

\subsection*{Dendrogram algorithms and evaluation of the partitions}
Using any/all the algorithms described above, a matrix of secondary distances is obtained from which dendrograms can be generated. Jerarca implements two well-known phylogenetic algorithms for this task, UPGMA and Neighbor-Joining. The user may run one or both algorithms.

From the dendrogram, partitions of the units into clusters can be obtained. Jerarca establishes partitions by scanning the dendrogram from the root to the external leaves. Starting from the root, each dichotomy in the tree (that increases the number of clusters) generates an alternative partition that can be evaluated. Given that the neighbor-joining method generates unrooted trees, the middle point of the tree is used as root \cite{22}. Jerarca implements two mathematically independent criteria in order to evaluate the community structure of a given partition. The first index is the well-known and broadly used modularity (Q) \cite{10}, which measures the distribution of within and between communities links in a certain partition compared to the expected number of connections that should exist given a specific degree distribution \cite{23}. The second index (called H) is based on the cumulative hypergeometric distribution of links, and derives from an index proposed in the paper that described UVCluster \cite{12}. The definition of H is as follows:

\begin{equation}
H=\displaystyle\sum\limits_{j=p}^{Min(M,n)} \frac{\binom{M}{j}{\binom{F-M}{n-j}}}{\binom{F}{n}}
\end{equation}

where $F$ is the maximum possible number of direct interactions in the whole network (for a network of k elements, $F = k (k-1)/2)$, $n$ is the number of direct interactions actually observed among the $k$ elements of the network, $M$ is the maximum possible number of intracluster direct interactions in a given partition and $p$ is the total number of direct intracluster interactions actually detected in that partition. The parameter $H$ measures the probability of obtaining by chance a given partition assuming a random distribution of intracluster and intercluster connections. The larger the value of $H$, the better ("more unexpected") the partition of the tree.

\subsection*{Output files}
Jerarca produces four types of output files (Figure \ref{fig:2}). Their names, automatically generated, include a reference to the algorithms and the evaluation criterion used (e. g. a typical name would be "Filename\_partitionH\_SCluster\_Upgma.txt"​). Moreover, the extension of a file specifies the content of the output:

\begin{enumerate}
\item Files with ".meg" extension contain the matrix of distances among units and the clusters obtained in the optimal partition of the dendrogram, according to either Q or H. This file can be directly imported into the software MEGA 4 \cite{24} for further analyses.
\item Files with ".att" extension contain the assignment of nodes to clusters in the best partition. These files are designed to be imported into Cytoscape (version 2.x) \cite{25} as attributes of the nodes (from the main Cytoscape menu: File - Import - Node attributes).
\item Files with a ".txt" extension save the best partition of the dendrogram obtained in text format. They include a description of the optimal partition: number of clusters, value of the index used and the assignment of nodes to each cluster.
\item Finally, the files with ".nwk" extension describe the dendrogram structure in standard Newick format, which can be read by virtually all programs that analyze trees, such as MEGA.
\end{enumerate}

\section*{RESULTS}
The speed of the programs has been tested in several benchmarks. Here we describe the results for three of them, consisting in an artificial and two real networks:
\begin{description}
\item[Benchmark A] \hfill \\
We prepared a synthetic graph of known community structure, in which 512 units were divided into 16 clusters of equal size. Within each cluster all units were initially fully connected (for a total of $(k^2-k)/2$ edges, being $k$ the number of units in a cluster). Then, we progressively "degraded" that structure by removing a certain percentage of edges and then randomly shuffling a number of edges among the units. The networks generated are a variation of the connected-caveman graphs defined by Watts \cite{26}.
\item[Benchmark B] \hfill \\
The proteins (nodes) that constitute 408 different protein complexes described in the yeast \textit{Saccharomyces cerevisiae} were obtained from the CYC2008 database (\href{http://wodaklab.org/cyc2008}{http://wodaklab.org/cyc2008}; \cite{27}). We then downloaded from the BioGRID database \cite{28} the protein-protein interactions (edges) characterized so far for all these proteins. The final graph contained 1604 nodes and 14171 edges.
\item[Benchmark C] \hfill \\
The complete set of protein-protein interactions (interactome) of \textit{S. cerevisiae} was obtained from BioGRID. These data generated a network formed by 5735 nodes (proteins) and 51134 edges (protein-protein interactions).
\end{description}

Benchmark A was specifically created for testing the quality of the optimal partitions computed by the algorithms implemented in Jerarca. We generated networks with progressive percentages of degradation. In this context, a percentage of degradation of, say, 10\%, means that first, 10\% of links were eliminated and, from the rest, 10\% shuffled among units. The shuffling process involves the random removal of an edge of the graph and the later addition of a new edge between two nodes, chosen also randomly. We previously suggested using a number of iterations equal to 10 times the number of units \cite{12}. Thus, for each of those networks, we ran 5000 iterations of Jerarca with the parameter \textit{all} for both the iterative and the tree algorithms. This means that 12 analyses ( = 3 iterative algorithmsx2 tree algorithmsx2 partition criteria) were performed for each network. With 0-30\% degradation, all algorithms recovered the original community structure of the network without errors. However, starting at 40\% degradation, slight errors in recovering the original community structure of the graph began to emerge, so we focused on this case. For each of the six dendrograms constructed by using the three iterative and the two tree algorithms, the optimal partitions given by the two evaluation indexes implemented in Jerarca (Q and H) were exactly the same. In all cases but one, a single unit of the network, different for each combination of programs, was misclassified. Only the combination of SCluster and UPGMA recovered the exact community structure of the original network. Significantly, this particular combination also obtained the highest Q and H values. This example shows that all the programs efficiently recover the original structure, even when it is quite cryptic (40\% degradation means that just about a third of the original links remain). On the other hand, it also shows the advantage of using when possible all the programs together, given that some may perform better than others.

We performed speed tests in a PC-compatible computer with an Intel Core 2 Quad Q8200 at 2.33 GHz and 4 GB of RAM, running Linux. The analyses of benchmark A were very fast. The 12 analyses per network described in the previous paragraph (5000 iterations/analysis) required just between 30 and 75 seconds. The least degraded ( = more compact) graphs, allow for the fastest analyses. To test the speed of the program in real networks of larger sizes, we used benchmarks B and C. For benchmark B (1604 nodes), 16000 iterations took about 3.25 hours when using the RCluster algorithm, while for UVCluster and SCluster the cost was 2 minutes and less than a minute respectively. This large difference is due to the fact that this network contains densely connected modules (each protein complex was much more tightly connected internally than with the rest of the network), a feature that favors the greedy strategies implemented in UVCluster and SCluster. For benchmark C (5735 nodes), 60000 iterations took 40 minutes with SCluster and about 3 hours with UVCluster. For RCluster, we estimated the analysis to require around 300 hours, so it was not performed in full.

In summary, the new algorithms implemented in Jerarca make possible to analyze large networks. As the times just detailed demonstrate, a single computer may easily cope with problems involving several thousands of units in a reasonable time, using both UVCluster and SCluster. Also, for networks with up to 1000 nodes, the user can test the three programs together, obtaining the results in minutes to a few hours.

\section*{DISCUSSION}
As the amount of biological information is rapidly increasing, one of the main goals in bioinformatics is the generation of fast programs able to deal with large datasets. For network analyses, the bottleneck of the iterative hierarchical clustering strategy is precisely that the clustering algorithm must be repeatedly used to generate a sufficiently large set of iterations as to be representative of the underlying structure of the graph. The second part of the analysis, the construction of the tree applying a phylogenetic algorithm is performed just once and therefore has little effect in the time complexity of the program. As already indicated in the Introduction, the applications of our UVCluster program were limited by the high amount of time needed for analyzing large networks. An optimization of the iterative clustering method implemented in that program was therefore mandatory. By setting certain restrictions (fixed \textit{AC}), we have qualitatively reduced the time complexity of the UVCluster algorithm. Traditionally limited to analyses below 1000 units, the current algorithm can cope with networks of several thousand units in a few hours. This allows analyzing some very interesting datasets, such as the whole interactome of the eukaryotic species \textit{Saccharomyces cerevisiae} (see benchmark C above).

A second significant advantage of Jerarca is that it also includes two novel algorithms, RCluster and SCluster, which provide alternative ways of computing the secondary distances between the nodes of the graph. RCluster randomly grows multiple clusters at the same time, avoiding the greedy agglomerative process implemented in UVcluster. However, the randomization process required makes the program slower than UVCluster. SCluster is just the opposite: it is the fastest and greediest of the three algorithms. In spite of its simplicity, its performance is also appropriate (See results for benchmark A above). Since Jerarca allows to execute several parallel analyses, we recommend to use the three iterative algorithms for networks with up to 1000 nodes. A complete analysis of such networks may require less than two hours (see Results). With larger networks, up to 10000 units, both UVCluster and SCluster can be used, the analyses with both programs requiring just a few hours. The inclusion of SCluster, which runs in $O(n log n)$ time, allows for the analyses of even larger networks. This may be of interest in fields such as the analysis of coexpression or gene interaction networks, in which the number of nodes (in those cases, corresponding to genes) may be in the tens of thousands. All these considerations obviously refer to analyses using a single computer. However, it is important to take into account that the programs can be very easily parallelized, given that the iterations can be divided into multiple processors and the results added together at the end of the computation.

In addition to UPGMA, already included in the original version of UVCluster, Jerarca also allows the alternative of building the trees using the neighbor-joining algorithm, probably the most frequently used algorithm to generate trees from a distance matrix. We suggest to obtain both trees (which is almost instantaneous), in order to evaluate the congruence of the results. An additional advantage of Jerarca respect to UVCluster refers to the determination of the optimal partitions of the graph according to two statistical parameters (Q and H). We added these options considering that the users may be often not only interested in obtaining a hierarchical representation of the network, but also in how the network can be divided into clusters or communities (see Introduction). The strategy used to obtain the partitions is in fact quite simple, given that the tree is just scanned from root to leaves. Therefore, the number of partitions examined is quite reduced (equal to the number of nodes n). More complex methods can be easily envisaged. For example, partitions could be generated at different distances from the root in different sections of the tree. However, although this option may potentially improve the likelihood of obtaining a better partition of the network, it is computationally much more expensive. We plan to explore this possibility in future versions of the suite. A final advantage of Jerarca is the set of outputs that it generates, which is much more complete than the one provided by our original UVCluster program. The possibility to directly export the data to powerful packages such as MEGA and Cytoscape will allow the users both to perform additional analyses that may complement those generated by Jerarca and to obtain sophisticated graphical representations of the results. All these advantages clearly make Jerarca a better tool to perform iterative clustering analyses of network data than our original UVCluster program.

The program, along with the source code is freely available under the GNU General Public License v3 at \href{http://jerarca.sourceforge.net}{http://jerarca.sourceforge.net}. The modular code structure of Jerarca permits easily including new features to the program. New algorithms, both iterative and for building the trees, as well as new indexes for extracting the optimal partition of the tree, can be easily added.

\section*{ACKNOWLEDGMENTS}
The authors would like to thank Vicente Arnau for his contributions in the development of the original UVCluster program, and Antonio Marco for his suggestions about how to implement the calculation of the H parameter.

\bibliography{Jerarca}

\end{document}